\pdfoutput=1

\documentclass[final,authoryear,3p,times,twocolumn]{elsarticle}

\usepackage{graphicx}
\usepackage{amsmath}
\usepackage{amssymb}
\usepackage{amsthm}
\usepackage{caption}
\usepackage{subcaption}

\biboptions{numbers,square}
\journal{Computer Physics Communications}

\begin{document}

\begin{frontmatter}

\title{Wave-Particle-Interaction in Kinetic Plasmas}

\author[nwu,jmu]{Cedric Schreiner\corref{cor}}
\ead{cschreiner@astro.uni-wuerzburg.de}
\cortext[cor]{Corresponding author}
\author[nwu]{Felix Spanier}

\address[nwu]{Center for Space Research, North-West University, 2520 Potchefstroom, South Africa}
\address[jmu]{Lehrstuhl f\"ur Astronomie, Universit\"at W\"urzburg, 97074 W\"urzburg, Germany}

\begin{abstract}
Resonant scattering of energetic protons off magnetic irregularities is the main process in cosmic ray diffusion.
The typical theoretical description uses Alfv\'en waves in the low frequency limit.
We demonstrate that the usage of Particle-in-Cell (PiC) simulations for particle scattering is feasible.
The simulation of plasma waves is performed with the relativistic electro-magnetic PiC code \textit{ACRONYM} and the tracks of test particles are evaluated in order to study particle diffusion.
Results for the low frequency limit are equivalent to those obtained with an MHD description, but only for high frequencies results can be obtained with reasonable effort.
PiC codes have the potential to be a useful tool to study particle diffusion in kinetic turbulence.
\end{abstract}


\end{frontmatter}


\section{Introduction}
\label{sec:introduction}

The transport of charged particles in the interstellar and interplanetary medium is governed by the scattering of those particles off magnetic irregularities.
The complete system of plasma and charged particles is highly nonlinear and the description of the processes is, therefore, very complicated. In order to gain some understanding the back reaction of energetic particles on the plasma is ignored in most cases.
The scattering itself may be described by different methods.
Since computers are readily available, different numerical models have been developed to describe particle scattering \citep{michalek_1996, qin_2002}.
Most methods assume random fluctuations in which particle tracks are followed.
The ansatz of {Lange et al.} \citep{lange_2013} assumed realistic turbulence.
While this limits the extent of the spectrum of magnetic fluctuations, the physics is described correctly.
We want to test, whether this ansatz may be used also for non-MHD plasmas.


\section{Theory}
\label{sec:theory}

For a statistical description of particle transport in magnetized plasmas we use the relativistic Vlasov equation.
We assume a collisionless plasma, where particle motion is only affected by the Lorentz force.
In order to examine the interaction of particles and waves analytically, we use the quasi-linear theory (QLT) which was first introduced by {Jokipii} \citep{jokipii_1966} and describes scattering processes of particles and waves as resonant interactions of the particle with the wave's electromagnetic fields.
Resonant scattering results in a significant change of the particle's direction of motion and occurs, if the resonance condition is met \citep{schlickeiser_1989}:
\begin{equation}
	k_\parallel ~ v_\parallel - \omega_0 + n ~ \Omega_\text{T} = 0.
	\label{res_con}
\end{equation}%
\noindent
Here $k_\parallel$ is the component of the wave vector parallel to the background magnetic field $B_\text{ext}$, $\omega_0$ is the wave's frequency, $\Omega_\text{T}$ is the (test) particle's cyclotron frequency and $n$ is the order of the resonance.
We will set $n=1$ throughout this article and hence consider only pure parallel waves.
The parallel component $v_\parallel$ of the (test) particle's velocity can also be written as the absolute value of its velocity multiplied by the cosine of the pitch angle: $v_\parallel = v_\text{T} \, \cos\theta = v_\text{T} \, \mu$.
\\
Time evolution of scattering processes can be analyzed using analytical functions for the maximum scattering amplitude after a period of time $t$, as provided by QLT \citep{lee_1974}:
\begin{eqnarray}
	\Delta \mu^\pm(t,\psi) =& \Omega_\text{T} ~ \sqrt{1 - \mu^2} ~ \dfrac{\delta B}{B_\text{ext}} \times
	\nonumber
	\\
	 &\dfrac{\cos{\psi} - \cos{((\pm k_\parallel \, v_\text{T} \, \mu - \Omega_\text{T}) ~ t + \psi)}}{\pm k_\parallel \, v_\text{T} \, \mu - \Omega_\text{T}}.
	\label{qlt}
\end{eqnarray}%
\noindent
The (+) sign applies to right handed polarized waves, whereas the (-) sign refers to left handed waves.
The amplitude of the wave's magnetic field is given by $\delta B$.
For a full representation of the scattering amplitude both the phase
\begin{equation}
	\psi^\pm(t) = \arctan{\left( \dfrac{\sin{((\pm k_\parallel \, v_\text{T} \, \mu - \Omega_\text{T}) ~ t})}{1 - \cos{((\pm k_\parallel \, v_\text{T} \, \mu - \Omega_\text{T}) ~ t)}} \right)},
	\label{qlt_phase}
\end{equation}%
\noindent
as well as the phase $\psi + \pi$ have to be considered.
Note that Eqs.~\eqref{qlt} and \eqref{qlt_phase} are given in the wave's rest frame and have to be transformed into the plasma rest frame by applying a Galilei transformation to $\mu$:
\begin{equation}
	\mu' = \mu - \omega_0 / (k_0 ~ v_\text{T}).
	\label{rest_frame}
\end{equation}%
\noindent
In Sect. \ref{sec:results} we will use the equations above to compare our numerical results to the predictions of QLT.
\\
In this article we discuss only transverse left handed waves with frequencies below the ion cyclotron frequency.
We also limit these waves to propagation parallel to the background magnetic field and we use the cold plasma approximation to derive their dispersion relation.
The general dispersion relation for the $L$-mode, which connects the wave number $k$ with the wave's frequency $\omega$, is given by \citep{canuto_1978}
\begin{equation}
	k_L = \pm \dfrac{\omega}{c} ~ \sqrt{1 - \frac{\omega_\text{p}^2}{(\omega + \Omega_\text{e}) ~ (\omega - \Omega_\text{p})}},
	\label{disp_rel}
\end{equation}%
\noindent
with the speed of light $c$, the plasma frequency $\omega_\text{p}$ and the cyclotron frequencies $\Omega_\text{p}$ and $\Omega_\text{e}$ of protons and electrons, respectively.
In its low frequency limit the dispersion relation Eq.~\eqref{disp_rel} includes the description of the Alfv\'en mode with propagation parallel to the background magnetic field.
Towards higher frequencies the wave mode is dispersive and reaches the cyclotron resonance at $\omega = \Omega_\text{p}$, where the low frequency branch of the $L$-mode ends.
The high frequency branch will not be considered in this article.


\section{Numerical approach}
\label{sec:numerics}

For our Particle-in-Cell (PiC) simulations we use the explicit second order PiC code \textit{ACRONYM}, which has been developed by our group and is fully relativistic, parallelized and three-dimensional \citep{kilian_2011}.
The PiC approach allows us to simulate the behavior of a thermal background plasma and a non-thermal population of test protons self-consistently.
This is one of the main differences to the simulations by {Lange et al.} \citep{lange_2013}, where a magnetized, turbulent background plasma is modeled magneto-hydrodynamically \citep[see][]{lange_2012_b}, whereas the test particles follow a kinetic approach.
\\
Furthermore PiC simulations allow the excitation of wave modes in the dispersive regime of the $L$-mode, while MHD simulations, such as the ones by \citep{lange_2013}, are limited to non-dispersive waves only.
\\
However, the advantages mentioned above come at the price of a higher need for computational resources, since PiC codes must resolve the microscopic scales of electron movement and are bound to very short time steps compared to typical proton time scales.
In order to reduce computational costs, we artificially reduce the proton mass by adjusting the ratio of proton to electron mass $m_\text{p} / m_\text{e}$, which is a common approach for PiC applications.
\\
\\
PiC codes tend to produce all kinds of physically possible wave modes in simulations of thermal plasmas.
Since we are interested in only one wave mode at a time, we introduce a mechanism to excite a single wave with specific wavenumber, frequency and polarization.
\\
We first choose a wave number $k_0$ in a way that an integer multiple of the wavelength fits into the periodic simulation box in $x$-direction.
Then we solve the dispersion relation Eq.~\eqref{disp_rel} for the wave to be excited, thus obtaining both wave number $k_0$ and frequency $\omega_0$.
By defining either the wave's electric or magnetic field strength $\delta E$ or $\delta B$, we set the transverse components of the wave's electromagnetic fields along the $x$-axis:
\begin{subequations}
	\begin{alignat}{2}
		\delta E_y &= \phantom{-} \delta E \, \sin{(k_0 \, x)}, & ~~~~~ \delta E_z &= - \delta E \, \cos{(k_0 \, x)}, \phantom{\frac{1}{1}}
		\label{excitation_e-field}
		\\
		\delta B_y &= - \frac{c \, k_0}{\omega_0} ~ \delta E_z, & ~~~~~ \delta B_z &= \phantom{-} \frac{c \, k_0}{\omega_0} ~ \delta E_y.
		\label{excitation_b-field}
	\end{alignat}
	\label{excitation}%
\end{subequations}
In the equations above $\delta E$ has to be defined; $\delta B$ then follows from Maxwell's equations (in cgs).
Note that the $y$- and $z$-component of each field are connected via the polarization properties of a left handed Alfv\'en wave.
\\
We initialize the simulation by allocating electromagnetic fields to each cell according to Eqs.~\eqref{excitation} and by loading a population of background protons and electrons with a thermal velocity distribution, where we superpose each particle's velocity with  an additional boost velocity $\delta\vec{v}$:
\begin{equation}
	\delta v_y = \frac{q_\text{s} \, c \, \delta E_z}{q_\text{s} \, B_\text{ext} - c \, m_\text{s} \, \omega_0},  ~~~ \delta v_z = \frac{- q_\text{s} \, c \, \delta E_y}{q_\text{s} \, B_\text{ext} - c \, m_\text{s} \, \omega_0}.
	\label{excitation_boost}
\end{equation}
Again, these equations only hold for left handed, circularly polarized waves and can be derived from the particle's equation of motion when only the Lorentz-force due to the electromagnetic fields of the wave and the magnetic background field (in $x$-direction) is considered.
The index s in Eqs.~\eqref{excitation_boost} denotes a specific species of particles -- electron or proton.
Charge $q_\text{s}$ and mass $m_\text{s}$ have to be adjusted accordingly.
\\
As a result of this procedure the simulation starts with a wave mode in a strongly excited state, which is far more intense than thermally excited waves and dominates the behavior of the plasma.
\\
\\
To analyze wave-particle-scattering we initialize an additional population of test protons which can be tracked individually and which hardly disturb the background plasma.
This is done by reducing the macro factor -- i.e. the number of physical particles represented by one numerical particle -- of the test protons to one, whereas typical background particles have macro factors of $10^8$.
For simplicity we initialize the test protons mono-energetically and distribute their direction of motion isotropically.
To study scattering processes we then simply track the changes in the cosine of the particles' pitch angles $\Delta\mu(t) = \mu(t) - \mu(t_0\!=\!0)$.


\section{Simulation setups}
\label{sec:simulations}

The simulations presented in this article aim at two points of interest.
First we analyze the effects of the artificial mass ratio to rule out the possibility of any unwanted or unphysical behavior in our simulations.
For that we adopt the setup used by {Lange et al.} \citep{lange_2013} -- non-dispersive Alfv\'en wave, mono-energetic test protons -- and try to reproduce the results as far as possible.
We also compare our results to the predictions of quasi-linear theory.
We use three setups with different mass ratios, which will be referred to as S\textrm{I}, S\textrm{II} and S\textrm{III}.
\\
Secondly we adapt setup S\textrm{III} and turn towards waves in the dispersive regime of the Alfv\'en mode.
We choose two wave modes with frequencies above the non-dispersive frequency range and study the scattering processes of test protons off these waves.
These two setups will be referred to as S\textrm{IV} and S\textrm{V}.
\\
For a further description of the different setups see table \ref{tab:phys_setup}, where different parameters of physical interest are presented.
\begin{table}[width=\linewidth]
	\caption{Physical parameters for different simulation setups.}
	\label{tab:phys_setup}
	\centering
	\begin{tabular*}{\linewidth}{@{\extracolsep{\fill} }c c c c c c c}
		\hline\hline
		Setup & S\textrm{I} & S\textrm{II} & S\textrm{III} & S\textrm{IV} & S\textrm{V}
		\\
		\hline\hline
		$m_\text{p} / m_\text{e}$ & $10.7$ & $21.4$ & $42.8$ & $42.8$ & $42.8$
		\\
		\hline
		$\omega_0 / \Omega_\text{p}$ & $0.108$ & $0.094$ & $0.103$ & $0.296$ & $0.500$
		\\
		\hline
		$\delta B / B_\text{ext}$ & $0.021$ & $0.028$ & $0.040$ & $0.045$ & $0.053$
		\\
		\hline
		$\mu_\text{res}$ & $0.56$ & $0.60$ & $0.66$ & $0.70$ & $0.70$
		\\
		\hline
		$v_\text{T} / c$ & $0.58$ & $0.49$ & $0.32$ & $0.16$ & $0.027$
		\\
		\hline
		$v_\text{A} / c$ & $0.051$ & $0.038$ & $0.027$ & $0.027$ & $0.027$
		\\
		\hline\hline
	\end{tabular*}
\end{table}
\\
The numerical setup, that is mainly the size of the simulation box, depends on the desired physical setting.
To optimize the setup regarding computing time, we use a simulation box of minimal size, which means that the long edge -- say, the $x$-axis -- comprises one wavelength of the excited wave mode.
For the other two edges of the box we choose lengths slightly larger than $2\, R_\text{L}$, with the Larmor-radius $R_\text{L}$ of a proton traveling at thermal velocity perpendicular to the background magnetic field $B_\text{ext}$.
This helps to avoid self-interaction of particles along their paths of gyration.
\\
Box sizes and Larmor-radii for the different setups can be found in table \ref{tab:num_setup}.
\begin{table}[width=\linewidth]
	\caption{Box sizes and Larmor-radii.}
	\label{tab:num_setup}
	\centering
	\begin{tabular*}{\linewidth}{@{\extracolsep{\fill} }l c c c c c}
		\hline\hline
		Setup & S\textrm{I} & S\textrm{II} & S\textrm{III} & S\textrm{IV} & S\textrm{V}
		\\
		\hline\hline
		$N_x$ [cells] & $3072$ & $5120$ & $6656$ & $2048$ & $1024$
		\\
		\hline
		$N_y \!=\! N_z$ [cells] & $56$ & $80$ & $112$ & $112$ & $112$
		\\
		\hline
		$R_\text{L}$ [cells] & $26.4$ & $37.3$ & $52.8$ & $52.8$ & $52.8$
		\\
		\hline\hline
	\end{tabular*}
\end{table} 
\\
\\
Other parameters of potential interest are given in the following:
We set our electron plasma frequency to $\omega_\text{p,e} = 5\cdot 10^8 \, \text{rad/s}$ and the thermal velocity of electrons to $v_\text{th,e} = 0.08 \, c$ so that the proton plasma frequency and the thermal velocity of protons become $\omega_\text{p,p} = \omega_\text{p,e} \cdot \sqrt{m_\text{e}/m_\text{p}}$ and $v_\text{th,p} = v_\text{th,e} \cdot \sqrt{m_\text{e}/m_\text{p}}$, respectively. The background magnetic field is $B_\text{ext} = 5 \, \text{G}$.


\section{Results}
\label{sec:results}

\begin{figure}[h!]
	\centering
	\begin{subfigure}{\hsize}
		\centering
		\includegraphics[width=\hsize]{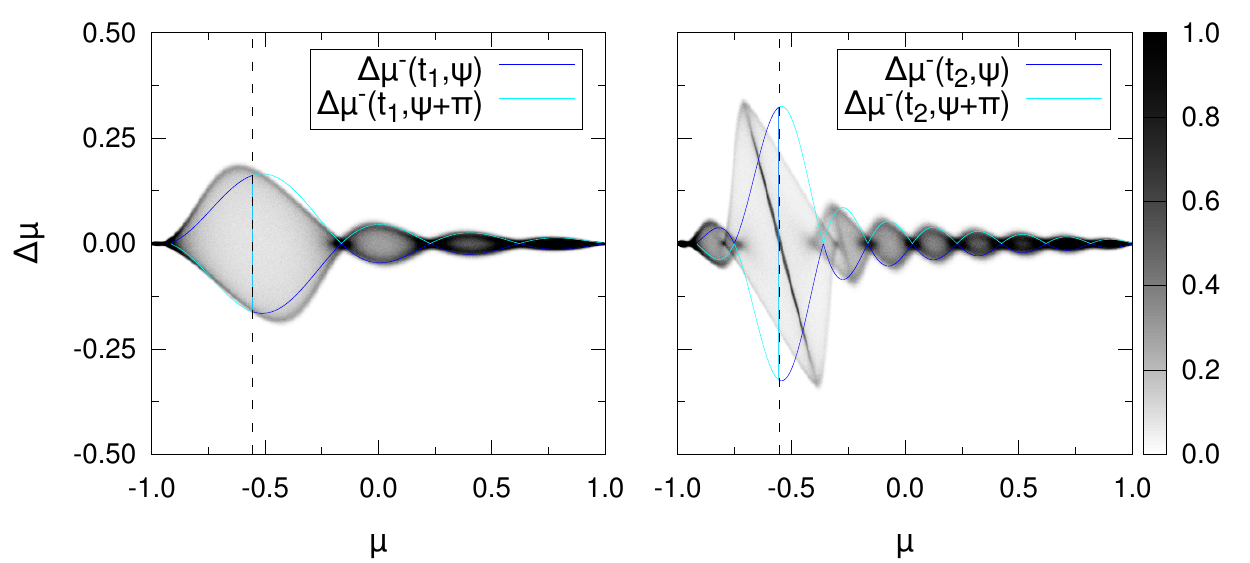}
		\caption{S\textrm{I}.}
		\label{fig:mpzume107}
	\end{subfigure}
	\\
	\begin{subfigure}{\hsize}
		\centering
		\includegraphics[width=\hsize]{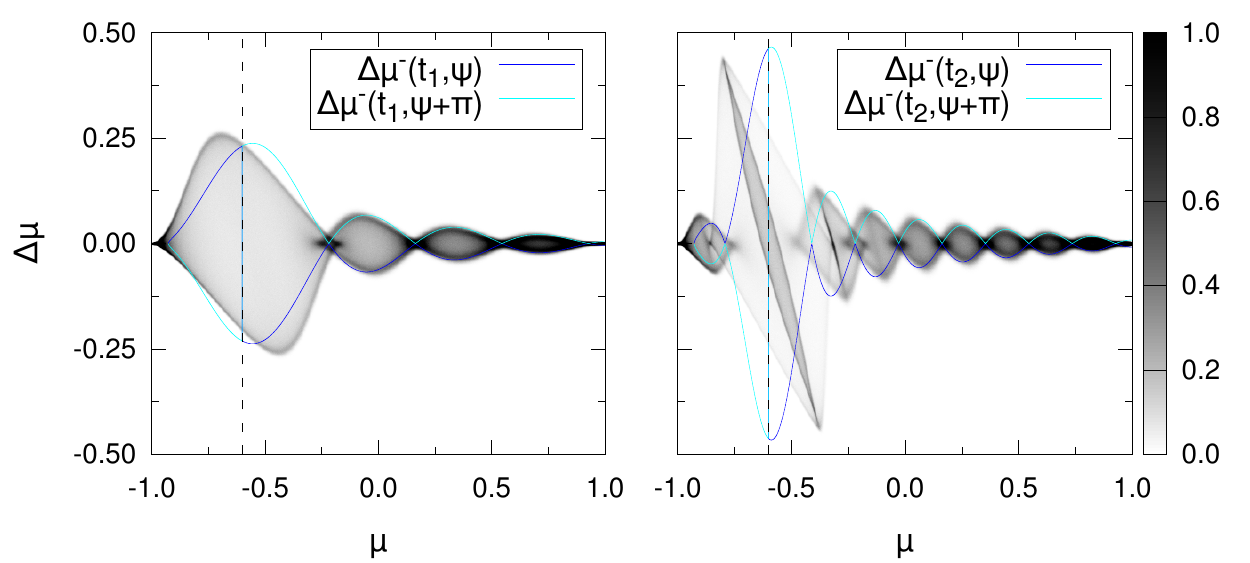}
		\caption{S\textrm{II}.}
		\label{fig:mpzume214}
	\end{subfigure}
	\\
	\begin{subfigure}{\hsize}
		\centering
		\includegraphics[width=\hsize]{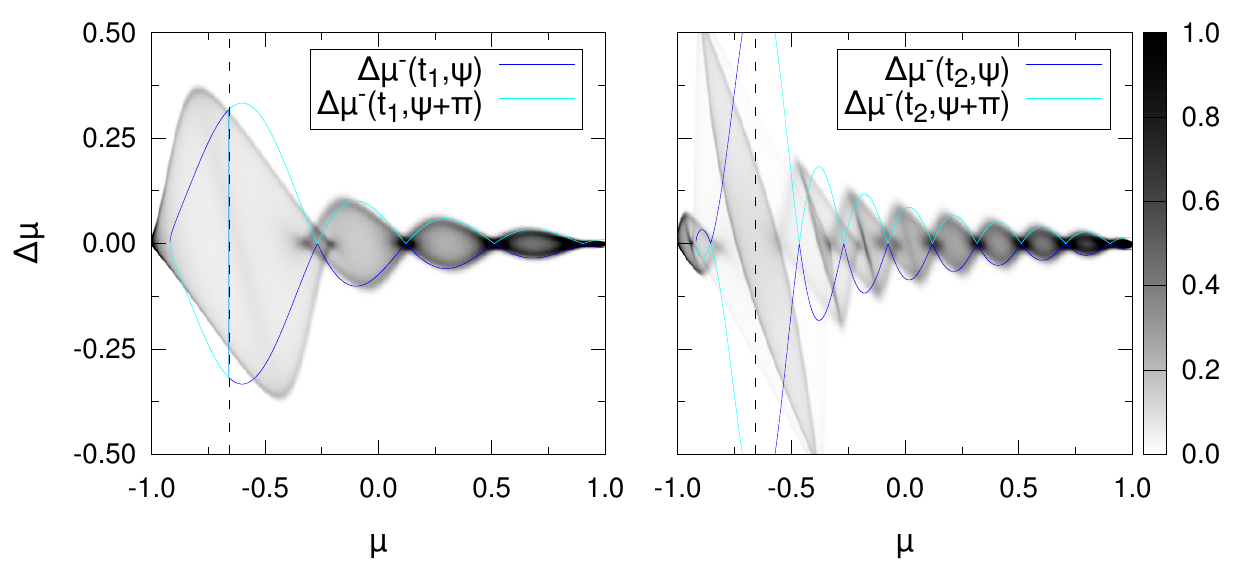}
		\caption{S\textrm{III}.}
		\label{fig:mpzume428}
	\end{subfigure}
	\caption{Pitch angle scatter plots for setups with different mass ratios $m_\text{p}/m_\text{e}$. The plots present the change $\Delta\mu$ of a particle's pitch angle at time $t$ depending on its initial $\mu$ at $t_0=0$, where $t_1$ and $t_2$ are intervals of two and four gyroperiods, respectively. Color coding indicates the number of particles at a given point $(\mu, \Delta\mu)$ normalized to $10^{-4}\cdot N_\text{T}$, where $N_\text{T}$ is the total number of test protons. The dashed lines show the position of the resonant pitch angle $\mu_\text{res}$ given by Eq.~\eqref{res_con}. The curves represent QLT predictions for the maximum scattering amplitude according to Eq.~\eqref{qlt}.}
	\label{fig:mass_ratio}
\end{figure}

\begin{figure}[h!]
	\centering
	\begin{subfigure}{\hsize}
		\centering
		\includegraphics[width=\hsize]{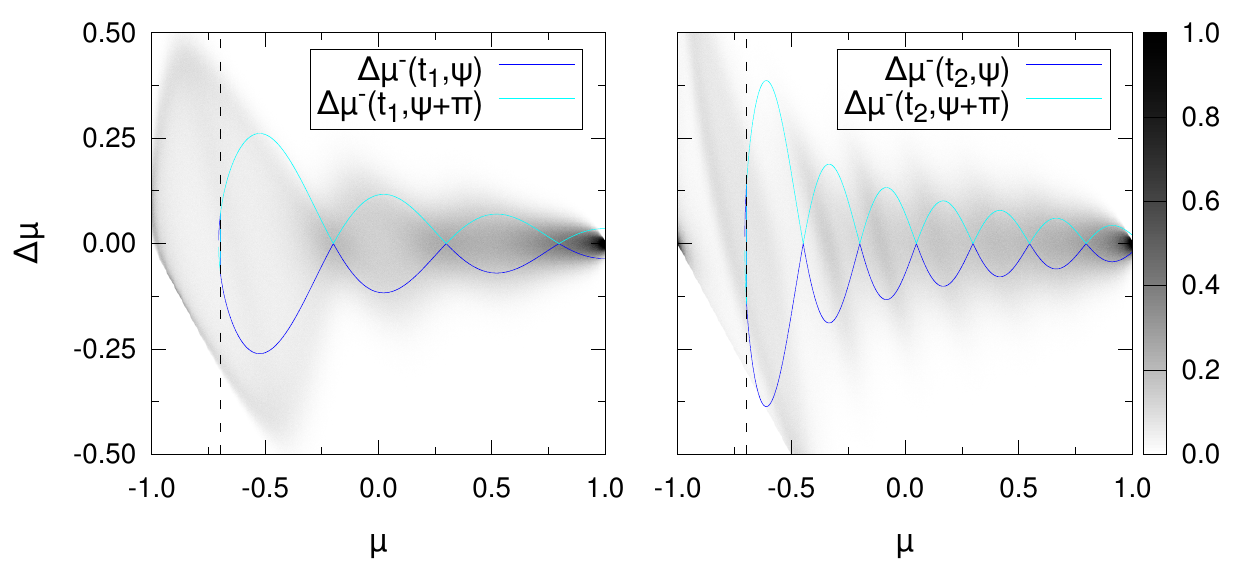}
		\caption{S\textrm{IV}.}
		\label{fig:disp029}
	\end{subfigure}
	\\
	\begin{subfigure}{\hsize}
		\centering
		\includegraphics[width=\hsize]{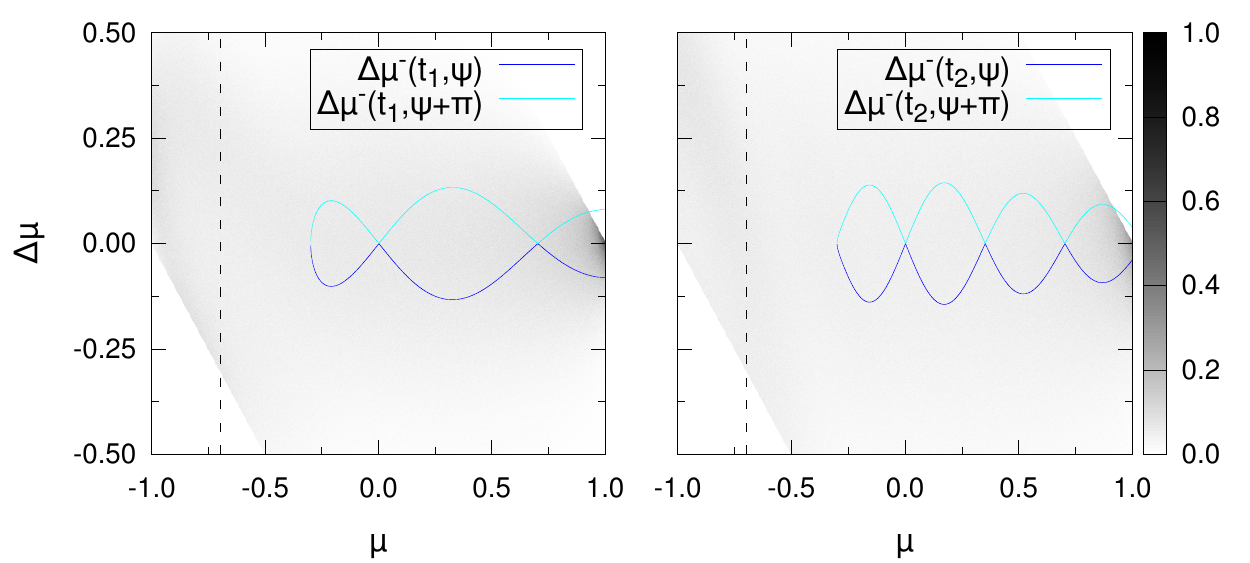}
		\caption{S\textrm{V}.}
		\label{fig:disp050}
	\end{subfigure}
	\caption{Pitch angle scatter plots for setups with different ratios $\omega / \Omega_\text{p}$. Again, $t_1$ and $t_2$ represent intervals of two and four gyroperiods, respectively. The color coding is analogous to Fig.~\ref{fig:mass_ratio}, the dashed line marks $\mu_\text{res}$, the curves show QLT predictions, which can only be obtained for a limited range of the initial $\mu$.}
	\label{fig:dispersive}
\end{figure}

In Fig.~\ref{fig:mass_ratio} we present pitch angle scatter plots for simulations S\textrm{I}, S\textrm{II} and S\textrm{III} analogous to the ones shown by {Lange et al.} \citep{lange_2013} for their MHD / test particle simulations.
We find resonant scattering, indicated by the main peaks in the scatter plots, as well as ballistic scattering, represented by the smaller peaks.
The peaks are tilted with respect to the analytic predictions due to the finite value of the perturbing magnetic field $\delta B$, which is neglected in QLT.
A first glance indicates that our simulations produce qualitatively equivalent results, regardless of the mass ratio.
These results agree with both QLT predictions and the results obtained by {Lange et al.} \citep{lange_2013}.
\\
Still there are obvious differences between the plots in Fig.~\ref{fig:mass_ratio}, which we will discuss in the next paragraphs.
Before doing so, we would like to point out that time scales are given in units of proton gyrations $T_\text{p} = 2\pi / \Omega_\text{p}$, with the proton cyclotron frequency $\Omega_\text{p}$.
This helps producing comparable plots, since the absolute time (in seconds) for the evolution of scattering processes differs with the mass ratio, whereas time evolution in units of $T_\text{p}$ is independent of $m_\text{p}/m_\text{e}$.
\\
Comparing the images on the left side of Fig.~\ref{fig:mass_ratio}, one finds that the scattering amplitude at time $t_1 = 2\, T_\text{p}$ varies with the simulation setup.
The reason for this is the ratio $\delta B / B_\text{ext}$ of the wave's magnetic field and the background field, which defines the maximum scattering amplitude according to Eq.~\eqref{qlt}.
Since this ratio is different in the three setups discussed here, we expect the maximum amplitudes $\Delta\mu_\text{max}$ to behave as $\Delta\mu_\text{max}^{\text{S}\textrm{I}} \!:\! \Delta\mu_\text{max}^{\text{S}\textrm{II}} \!:\!\Delta\mu_\text{max}^{\text{S}\textrm{III}} = \delta B^{\text{S}\textrm{I}} \!:\! \delta B^{\text{S}\textrm{II}} \!:\! \delta B^{\text{S}\textrm{III}}$.
In fact, this behavior can be found in the scatter plots.
\\
The images on the right hand side of Fig.~\ref{fig:mass_ratio}, taken at time $t_2 = 4\, T_\text{p}$, show that time evolution seems to proceed at different speeds, depending on the setup.
This becomes especially clear when looking at the substructure of the resonance peaks, which is only a thin line in Fig.~\ref{fig:mpzume107} and broadens in Figs.~\ref{fig:mpzume214} and \ref{fig:mpzume428}, meaning that the evolution of scattering processes is most advanced in simulation S\textrm{III}.
We explain this discrepancy as follows:
Although our timescale measured in $T_\text{p}$ is independent of $m_\text{p} / m_\text{e}$, we did not take relativistic corrections into account.
Test protons are traveling at relativistic speeds, so their cyclotron frequency is decreased.
Since the test particles in S\textrm{I} are faster than those in S\textrm{II}, scattering processes develop slower in this simulation, according to Eq.~\eqref{qlt}.
The same holds for a comparison of S\textrm{II} and S\textrm{III}.
\\
Having outlined the main reasons for the different appearance of the scatter plots for setups S\textrm{I}, S\textrm{II} and S\textrm{III}, we find that there are no features present, which can not be explained by the set of input parameters or the form of presenting our data.
We, therefore, conclude that the artificial mass ratio does not cause any unforeseen or unphysical artifacts and that our results can be extrapolated to the natural mass ratio.
\\
\\
Our second set of simulations, S\textrm{IV} and S\textrm{V}, provides the results given in Fig.~\ref{fig:dispersive}.
Since the scatter plots in Figs.~\ref{fig:disp029} and \ref{fig:disp050} differ severely, we will first discuss the results of S\textrm{IV} and treat those of S\textrm{V} separately.
\\
Fig.~\ref{fig:disp029} basically shows the expected structures:
We find resonant scattering at the position predicted by the resonance condition Eq.~\eqref{res_con} and ballistic scattering, as indicated by the smaller peaks.
The analytical curves agree with the data in the range where QLT predictions are possible, but do not yield a correct representation of the full resonance peak.
Here our simulation shows that resonant scattering off waves in the dispersive regime can occur, although analytic theory can not give the full information about this process.
\\
On the left side of Fig.~\ref{fig:disp029} it can be seen that the peak structures are not sharply confined, like the ones in Fig.~\ref{fig:mpzume428} are.
This becomes even more evident as time proceeds.
We explain these diffuse edges of the peaks by particle scattering off random perturbations in the thermal background plasma.
Compared to the previously discussed simulations the effect is more significant in simulation S\textrm{IV}, since the test protons in this setup are slower, with a speed comparable to that of particles in the high energy tail of the thermal population.
Therefore test protons in S\textrm{IV} are affected by thermal effects more strongly than those in previous simulations.
\\
Regarding the results of S\textrm{V} in Fig.~\ref{fig:disp050} one hardly finds any structure.
Although the test particles' velocities and the parameters of the excited wave meet the resonance condition, there seems to be only a diffuse background of scattered particles.
The reason for this cannot be derived from the theory of resonant wave-particle-scattering, but is found in the properties of the excited wave mode itself:
The wave's frequency is close enough to the proton cyclotron resonance at $\Omega_\text{p}$ so cyclotron damping plays an important role.
A more detailed analysis of simulation S\textrm{V} has shown, that the excited wave is damped strongly and that it is dissipated almost completely during the first two proton cyclotron timescales $T_\text{p}$, meaning that at $t_1 = 2\, T_\text{p}$ no dominant wave is left in the background plasma.
Therefore test particles only scatter off random fluctuations and no resonant scattering can be achieved.
Although we have underestimated cyclotron damping at the time we planned the simulation setup, we are still convinced that resonant scattering off dispersive waves is possible even at wave frequencies similar to that in our simulation S\textrm{V} (see the discussion in Sect. \ref{sec:discussion}).

\section{Quality of results and comparison to MHD}
\label{sec:comparison}

To evaluate the quality of our results and to point out potential benefits or drawbacks of the PiC approach, we compare our simulations to those in {Lange et al.} \citep{lange_2013}.
\\
\\
Before doing so, we would like to briefly discuss a key concern, which is often addressed when PiC results are presented.
PiC simulations show a typical noise in the electromagnetic fields caused by a relatively small number of macro particles \citep[see e.g.][]{birdsall_2005}, which often affects the quality of PiC results.
Although the fields in the background plasma of our simulations are subject to PiC-specific noise effects, most of the results shown in this article are hardly influenced by this noise.
We come to this conclusion by analyzing the setup for our simulations:
\\
A strongly amplified wave mode is excited at the beginning of the simulation.
The excited wave dominates the fluctuations in the background plasma and the amplitudes of its electric and magnetic fields exceed the field strengths of the ``natural'' background, including noise. 
Test protons, which are the only population of particles relevant for our further analysis, mainly interact with the magnetic field of the excited wave and thus will rarely be influenced by noise.
\\
This argumentation holds at least for simulations S\textrm{I}, S\textrm{II} and S\textrm{III}.
As discussed in Sect. \ref{sec:results}, random fluctuations in the background plasma become relevant for the behavior of test protons in simulation S\textrm{IV} and even dominate the transport characteristics of those in S\textrm{V}.
It is hard to distinguish, if the diffuse scattering, which can be seen in Fig.~\ref{fig:disp029}, has its origin in noise effects or actual physical perturbations.
Since the results shown in this article are only meant as a proof of concept and computational resources for our simulations were limited, we did not conduct any further studies.
A simple, but computationally expensive way to investigate the influence of noise would be to re-run the simulation with different numbers of background particles per cell.
\\
\\
Basically, the quality of results in form of scatter plots, such as the ones shown in \citep{lange_2013} or in this article, depends on test particle statistics and implementation.
With given electromagnetic fields in the background plasma -- either in a PiC or MHD simulation -- the test particles react to those fields, which means their equation of motion is given by the Lorentz force.
Within the numerical implementation, field strengths have to be interpolated to the position of the particle and each particle has to be moved accordingly.
Both the \textit{Gismo} MHD code used by \citep{lange_2013} and the \textit{ACRONYM} PiC code \citep[see][]{kilian_2011} employ the so-called \textit{Boris-Push} \citep{boris_1970} to model a physically correct particle movement.
Therefore, test particle behavior should be equivalent in both kinds of simulations.
\\
Although the algorithm is the same in both codes, there is still a major difference.
At the beginning of each \textit{Gismo} simulation run, the optimal length $\Delta t$ of the numerical timestep is calculated as described in \citep[][Sect. 2.1]{lange_2013}, whereas the PiC algorithm requires a fixed length given by the CFL-condition.
As a result of these two different constraints for the maximum length of a timestep, $\Delta t_\text{MHD}$ is much larger than $\Delta t_\text{PiC}$.
Whereas a PiC code might be slightly more accurate in describing the trajectory of test protons, an MHD code is simply more efficient in terms of computational effort.
\\
\\
For the statistical significance of the results in \citep{lange_2013} and in this article, only the number of test protons per simulation matters.
Differences in the quality of the scatter plots become especially clear when one compares the right hand side of Fig.~\ref{fig:mass_ratio} in this article to \citep[][Fig.~1]{lange_2013}:
Whereas both figures show the overall shape of the scattering amplitudes, the resolution of the substructure differs noticeably.
The plots presented in this article clearly benefit from the larger number of test particles and are able to reveal much more detail.
\\
A large number of test particles is easily possible in a PiC simulation, since -- because of the even larger number of background particles -- the overall computational effort is only slightly increased in respect to a standard PiC simulation.
In the case of the simulations presented in this article, we include one test proton per grid cell, leading to $9\cdot 10^6 - 8\cdot 10^7$ test protons, depending on the setup (see Table \ref{tab:num_setup}).
Still, this immense number of test particles contributes only $\sim 6\%$ to the computational cost of the whole simulation.
\\
For hybrid MHD codes, such as \textit{Gismo}, the situation is different, because the performance depends sensitively on the number of test particles.
\\
\\
To end the comparison between PiC and MHD approach, we would like to stress that in those cases, where MHD is applicable, an MHD simulation is way more efficient.
To produce scatter plots of comparable quality within an equivalent physical setting, such as those of \citep[][Fig.~1]{lange_2013} and Fig.~\ref{fig:mass_ratio} in this article, the computational costs (in CPU-hours) for a PiC or an MHD simulation differ by approximately one to two orders of magnitude, depending on the mass ratio used in the PiC simulation.
Since the computational effort in the setups presented in this article scales with $(m_\text{p}/m_\text{e})^{2.5}$ for a given normalized wave frequency $\omega_0 / \Omega_\text{p}$, it is clear that simulations with mass ratios above $m_\text{p}/m_\text{e} \approx 50$ are not feasible.
The same holds for our simulations in the dispersive regime of the $L$-mode.
\\
However, in all cases which can not be covered by MHD, PiC is essential due to the lack of alternative methods.


\section{Discussion and conclusions}
\label{sec:discussion}

Our results show that PiC simulations are suitable to model wave-particle-interactions in kinetic plasmas.
As a very important outcome of our analysis, we would like to stress that resonant interactions are independent of the artificial mass ratio used PiC simulations.
In the low frequency regime of dispersionless Alfv\'en waves we are able to reproduce the MHD results of {Lange et al.} \citep{lange_2013} and find scattering amplitudes in accordance with QLT predictions.
These results do not allow any new insights regarding the physics of the problem, but are useful and necessary to validate our simulations and to investigate the effects of the artificial mass ratio.
\\
\\
The more interesting test case presented in this article is the scattering of particles off dispersive waves, since this cannot be done in MHD simulations and the QLT does not yield the full information about the problem.
Our simulation S\textrm{IV} indicates that resonant scattering off waves in the dispersive regime shows the same characteristic behavior regarding scatter plots, which can be found in the first three simulations S\textrm{I}, S\textrm{II} and S\textrm{III}.
Yet, the frequency range accessible has an upper limit, since the waves are damped near the proton cyclotron frequency.
\\
We propose to solve the problem of wave dissipation by changing the mechanism, which excites the wave.
Instead of initializing fields and particle velocities only at the beginning of the simulation, a constant input of energy in form of electromagnetic fields matching the wave's polarization and propagation properties should be able to drive the wave mode over a period of time sufficiently long to establish resonant scattering.
\\
\\
Although PiC simulations of wave-particle-scattering within the dispersive regime of a wave mode appear to be promising and might lead to new results which can not be obtained from MHD simulations, they are still very demanding in terms of computational resources.
The simulations S\textrm{IV} and S\textrm{V} presented in this article consumed several tens of thousands of CPU-hours each.
Therefore, such simulations can only be conducted, if plenty of computing time is available and if the process to be modeled allows for a rather low mass ratio.
Also, the number of background particles should be kept low -- so noise in the background plasma might become an issue.
\\
\\
In conclusion, it appears to us that a full scale application of PiC simulations describing the scattering of protons off of low frequency waves in a truly physical setting -- i.e. with realistic solar wind parameters -- is not in sight.
Although simple conceptual studies are feasible and reasonable, we see another potential benefit of PiC simulations in a related topic:
Instead of using test protons and waves in the low frequency regime of the $L$-mode, we propose to probe the scattering of relativistic test electrons off Whistler waves.
\\
Since electrons are lighter and faster than protons, the relevant time scales become shorter and less timesteps have to be carried out to describe scattering processes.
Also, Whistler waves have higher frequencies and shorter wavelengths than low frequency (Alfv\'en) waves, meaning that the simulation size can be decreased.
It is even possible -- or advisable -- to increase the mass ratio to its natural value, since protons should not contribute significantly to the transport characteristics of electrons.
The protons' Larmor radii do not have to be resolved within the simulation box, because with heavier protons and thus slower proton motion, their gyromotion becomes negligible on electron time scales.
Therefore, the simulation size is defined only by the wavelength and the electron Larmor radius.
\\
As a crude estimate for a typical problem size, the setup of simulation S\textrm{V} gives a good clue.
Relative to S\textrm{V} the number of timesteps can be reduced by a factor of $m_\text{p}/m_\text{e}=42.8$, since this is also the ratio of electron to proton cyclotron frequency.
This also suggests that the computational cost for the simulation will come close to that of a typical MHD simulation by {Lange et al.} \citep{lange_2013}, as discussed at the end of Sect. \ref{sec:comparison}.
\\
We are planning to carry out such a simulation in a follow-up project, since we are of the opinion that simulations of that kind are feasible and results can be obtained in a way similar to the procedure described in this article.
\\
Furthermore, we see PiC simulations as a promising approach to study particle diffusion in kinetic turbulence.
Recent findings by {Howes et al.} \citep{howes_2008} and {Che et al.} \citep{che_2014} state that magnetic turbulence can be reproduced in kinetic simulations, using either a gyrokinetic code \citep{howes_2008} or a 2.5-dimensional PiC code \citep{che_2014}.
Although we appreciate the results obtained by \citep{che_2014}, we suggest fully three-dimensional PiC simulations, because the tracking of particle trajectories along magnetic field lines can only lead to realistic results when the magnetic field is fully developed in all spatial dimensions.

\section*{Acknowledgement}
The authors gratefully acknowledge the Gauss Centre for Supercomputing e.V. (www.gauss-centre.eu) for funding this project by providing computing time on the GCS Supercomputer SuperMUC at Leibniz Supercomputing Centre (LRZ, www.lrz.de).
CS would like to thank Alex Ivascenko for helpful discussion and information given about the performance of the \textit{Gismo} MHD code.

\bibliography{bib}

\end{document}